\documentclass[%
 reprint,
 amsmath,amssymb,
 aps,
]{revtex4-1}
\usepackage{color}
\usepackage{xcolor}
\usepackage{textcomp}
\usepackage{graphicx}
\usepackage{dcolumn}
\usepackage{bm}

\definecolor{air}{rgb}{0, 0.2704, 0.5376}
\definecolor{airr}{rgb}{0.2, 0.07, 0.48}
\definecolor{prusia}{rgb}{0.0, 0.19, 0.33}
\definecolor{oxford2}{rgb}{0.0, 0.169, 0.336}
\definecolor{oxford}{rgb}{0.0, 0.2197, 0.4368}
\definecolor{math}{rgb}{0, 0.2704, 0.5376}
\definecolor{redmath}{rgb}{0.0, 0.2197, 0.4368}
\definecolor{veneciano}{rgb}{0, 0.2704, 0.5376}
\definecolor{spin1}{rgb}{0.7, 0.3, 0.}
\definecolor{spin2}{rgb}{0.3, 0.7, 0.}
\definecolor{oxfordfootnote}{rgb}{0.0, 0.44852564102564102564102564102564, 0.25876923076923076923076923076923}
\definecolor{oxfordapendix}{rgb}{0., 0.17, 0.38}

\begin{document}

\title{ \color{prusia}{Quantum simulation of the integer factorization problem: \\ Bell states in a Penning trap.}}

\author{\color{oxford}{Jose Luis Rosales}}
 \email{\color{oxfordfootnote}{Jose.Rosales@fi.upm.es}}
\author{\color{oxford}{Vicente Martin}}
 \email{\color{oxfordfootnote}{Vicente@fi.upm.es}}

\affiliation{%
 \color{oxford}{Center for Computational Simulation (Madrid) \\
 DLSIIS ETS Ingenieros Inform\'{a}ticos, Universidad Polit\'{e}cnica de Madrid,\\
Campus Montegancedo, E28660 Madrid.}
}

\date{\color{oxfordfootnote}{\today}}

\begin{abstract}

\color{oxford}{
The arithmetic problem of factoring an integer $N$ can be translated into the physics of a quantum device, a result that supports P\'{o}lya's and Hilbert's conjecture to prove Riemann's hypothesis. The energies of this system, being univocally related to the factors of $N$, are the eigenvalues of a bounded Hamiltonian.  Here we solve the quantum conditions and show that the histogram of the discrete energies, provided by the spectrum of the system, should be interpreted in number theory as the relative probability for a prime to be a factor candidate of $N$. This is equivalent to a quantum sieve that is demonstrated to require only $ o(\log \sqrt N)^3$ energy measurements to solve the problem,  recovering Shor's complexity result. Hence, the outcome can be seen as a probability map that a pair of primes solve the given factorization problem. Furthermore, we show that a possible embodiment of this  quantum simulator corresponds to two entangled particles in a Penning trap.  The possibility to build the simulator experimentally is studied in detail. The results show that factoring numbers, many orders of magnitude larger than those computed with experimentally available quantum computers, is achievable using typical parameters in Penning traps.}

\begin{description}
\item[PACS numbers] \color{oxfordfootnote}{03.67.Ac,03.67.Lx,02.10.De}
\end{description}
\end{abstract}

\maketitle

\color{prusia}{
\section{Introduction.}
}  
\color{oxford}{A classical computer, using the best factoring algorithms known at present~\cite{Pomerance}, requires a number of steps that grows exponentially with $l$, the number of digits of $N$, the integer to factorize.  Indeed, for large numbers the intractability of the factorization problem  underlies the security of  cryptographic algorithms like RSA or Diffie-Hellman~\cite{Koblitz}.

However, following the principles of quantum mechanics, a computer will solve the problem in polynomial time when it is programmed with  Shor's algorithm~\cite{shor}. The exponential speed up  is due to the quantum interference of probability amplitudes during  unitary evolution of the states in the appropriate Hilbert space, a property of the quantum Fourier transform on which the algorithm is based. This has been demonstrated experimentally, showing the soundness of the approach. For example, the number  $15$ was successfully factorized with this algorithm using a molecule serving as a seven qubits quantum computer~\cite{Chuangetal}. However,  even for this small $N$, preparing a fully programmable quantum computer is  still a  significant experimental challenge because it requires coherent control over many qubits. The scalability of this approach is complicated. It requires to employ sophisticated quantum error correction codes, ultimately needing millions of physical qubits to implement the thousands of logical qubits required to factor numbers in the range of $\sim 1$ kbit, the typical RSA-size integers used today.  Other approaches, like resorting to the preparation and measurement of complex states embodying properties of the primes~\cite{latorre14} is also difficult and has not been  realized up to now.

On the other hand, we have recently  proposed an equivalent formulation of the factorization problem $N=xy$ where the factor $x$ is replaced by the value of a  function $E(x)$ which is defined for the prime $x\leq\sqrt{N}$ such that there is another prime $y=N/x$~\cite{rosales-martin}. Thus, while Shor's algorithm reproduces the outputs of an arithmetic function that is periodic, module $N$, to find $x\leq \sqrt{N}$, the new formulation is tailored to find the probability distribution of $E(x)$ within a finite ensemble of prime numbers.
 Since every possible  factor of $N$ belongs to this set, we called it the factorization ensemble. Moreover, owing to the statistical properties of $E(x)$, a probability for a given $x$ to be a factor of $N$ could be inferred.  The new formulation can be translated to the physics of a two dimensional system  with bounded trajectories
that,  using semiclassical quantization, could be interpreted as the classical counterpart of a quantum factoring simulator if $E$ is identified with the energy. This approach will be correct for very large $N$ which, indeed, is the  more relevant and practical case.

The paper is organized as follows. First, in section II, we revisit, for very large numbers $N$, the new   formulation of the factorization problem. Then, we introduce the arithmetic function that will correspond to the Hamiltonian of the quantum simulator. Also, in this section, in order to put things into a context, a short review of number theory is given to let the reader understand how the quantum theory of the factorization problem should indeed be connected to Riemann's hypothesis. The quantum conditions for the stationary states of the simulator are solved in section III, obtaining a discrete quantum spectrum of energies when the state is prepared with the boundary conditions needed for the primes in the factorization ensemble. A  physical realization of the quantum state is presented  in Section IV for two entangled particles in a Penning trap. The spectrum of the arithmetic function $E$ is predicted to correspond to the magnetron energies of this state, it requires to program the number $N$ in terms if the physical parameters of the trap. This follows with  our conclusions. To facilitate the reading, the more cumbersome calculations are presented in the Appendix.

\bigskip
\color{prusia}{

\section{The Hamiltonian of the Quantum Simulator.}
}

The prime-ordering of $x$, in the list of all primes, is the arithmetic function $\pi(x)$. It is also known as the prime counting function less than a given magnitude $x$. Thus, for instance $\pi(3)=2$, $\pi(5)=3$, $\pi(101)=26$, etc. Each prime $x$ is univocally  given  by its prime counting function $\pi(x)$. Thus, the solution of the problem of factoring a number $N$, is either to obtain the value of $x$ from some algorithm or, equivalently, to determine its prime-ordering function $\pi(x)$. For instance  $26$ has the factors $x=2\leftrightarrow\pi(2)=1$ and $x=13\leftrightarrow\pi(13)=6$. This means that, given  the solution of factoring $26$, a bijection between both pairs of numbers can be determined univocally. In order to investigate the application \[\displaywidth=1.0\linewidth \mathcal{C}_{\mathbf{\mathfrak{F}}}(x\cdot y)\mapsto(\pi(x),\pi(y)),\] and to research the algorithm complexity of finding these factors of $N$, we have recently introduced the factorization ensemble~\cite{rosales-martin}, to which a given $N$ should belong.  It is defined as the set of all primes $x_k$ and $y_k$ such that their product yields numbers $N_k$, in a vicinity of $N$, with the property $j=\pi( \sqrt{N_k)}=\pi( \sqrt{N})$.
\begin{equation}
\label{FactorizationEnsemble}
\color{redmath}{\mathbf{\mathbf{\mathfrak{F}}}(j)=\left\{x_k, y_k \in\mathfrak{P} \mathbin|  N_k=x_k\cdotp y_k \: \wedge \:   \pi(\sqrt{N_k})=j\right\}}.
\end{equation}
For instance, the numbers $26=2\cdot 13$  and $25=5\cdot 5$  belong to the factorization ensemble $\mathbf{\mathfrak{F}}(3)$, since in both cases $j=\pi(\sqrt{26})=\pi(\sqrt{25})=3$.

Now, in order to characterize numerically the existence of the factorization equivalence class between the  prime factors and the prime ordering of the factors of a given $N_k=x_k y_k$  in the ensemble, we introduced also in~\cite{rosales-martin} the arithmetic function
\begin{equation}
\label{Energies}
\color{redmath}{E_k=\pi(x_k)\pi(y_k)/j^2}.
\end{equation}
Since $E_k\leftrightarrow (x_k, y_k)$ univocally,  the solution $(x,N)$ of the factorization problem can  be rewritten as the pair $(E,N)$.

Again, to clarify this definition, let us compute $E_1$ and $E_2$ for $N_1=26$ and $N_2=25$ ( $N_1, N_2 \in \mathbf{\mathfrak{F}}(3)$):
\[
\displaywidth=0.5\linewidth  E(26)=1\cdot 6/3^2=2/3, \;\;\;
E(25)=3\cdot 3/3^2= 1.
\]
Thus, the numbers $(2,26)\leftrightarrow (2/3,26)$ and $(5,25)\leftrightarrow (1,25)$ are the solutions of their respective factorization problem in terms of the primes and of the prime ordering functions of the respective solutions. The prime counting function has been the object of extensive research. Using Euler's identity, the complex function \[\zeta(s)=\sum_{1}^{\infty}\frac{1}{n^s}=\prod_{p\in\mathfrak{P}}\frac{1}{1-1/p^s},\]  can be set up with the help of the primes and, thus, characterize $\pi(x)$. As a matter of fact, after the work of Riemann~\cite{Riemann}, a Fourier  series, written in terms of  the zeroes of  $\zeta(s)$ on the critical line in the complex plane, $\rho_k=1/2+i\sigma_k$, would  exactly determine $\pi(x)$. Indeed,  the statement that every $\sigma_k$ is a real number, known as the Riemann hypothesis, is the cornerstone of number theory. On the other hand, P\'{o}lya and Hilbert conjectured that the hypothesis will be true if the  $\sigma_k$'s were eigenvalues of an Hermitian operator  or, in modern terms, if this operator has the matrix form of a quantum Hamiltonian in Hilbert space (see e.g.~\cite{schu} for a review of relation of the conjecture with physics). This line of research has not yet obtained the reward of success, mainly because, in order to achieve the quantization  of the states of the Hamiltonian, it must be necessarily bounded and, unfortunately,  no such a bound has been found without assuming {\it ad hoc} conditions depending on the proposal.

Moreover, since the distribution of the primes and the zeroes of the $\zeta(s)$ function are closely related, we can recast  P\'{o}lya's and Hilbert's conjecture in terms of the existence of a quantum system that obtain, as energy eigenvalues, the distribution of the primes $\pi(x)$ for $x\leq\sqrt{N}$ (equivalently the function $E$ defined above in the factorization problem). This imposes a natural bound to the primes and, therefore, the quantization could be obtained without  any need for additional assumptions.  This paper tries to follow this program to support these ideas.

The prime number theorem states that $\pi(x)\sim x/\log x$ for large $x$, then  implying that, for $x=o(\sqrt{N})$, and very large $N$, $E$ scales  logarithmically with the factor $x_k$. Hence, linear increments in $E$ require large exponential increments in $x$.

Let us return to our original considerations on how to find a quantum system that relates the factors with the eigenvalues of the arithmetic function $E$ defined above. From these definitions, the set $(E_k,N_k)$  can be calculated. In Fig.~\ref{fig:spectrum} we present the results for $j=10000$,  showing  the typical band structure of a quantum spectrum. This result can not be directly explained from number theory but it can be readily understood when considering that those points correspond to the measurements of the energy observable "$E$" of the quantum simulator proposed in~\cite{rosales-martin}.
\begin{figure}[hbtp]
\centering
\includegraphics[scale=0.43]{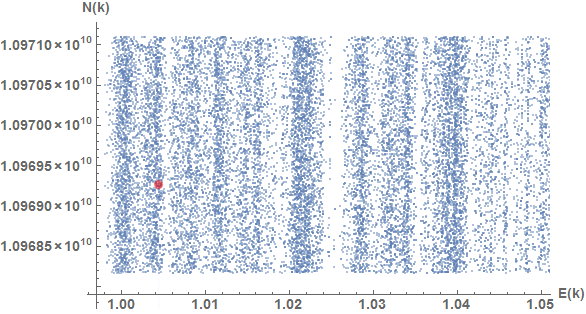}
\caption{\color{oxfordfootnote}{A plot of  the pairs of values $(E_k,N_k)$ in the factorization ensemble  $\mathbf{\mathfrak{F}}(10000)$ showing the typical band spectrum of a quantum system. This cannot be expected from number theory alone.  As an example, the point $N=10969262131 =
47297\cdotp 231923$, $E=1.00441815$ is marked with a larger dot.}}
\label{fig:spectrum}
\end{figure}

Now we can define the variables $q=(\pi(x)+\pi(y))/2j$ and $p=(\pi(y)-\pi(x))/2j$ for $x<y$. Then Eq.~(\ref{Energies}) is transformed into the inverted harmonic oscillator Hamiltonian function
\begin{equation}\label{invertedharmonic1}
\color{redmath}{E(x,y)=\pi(x)\pi(y)/j^2 \leftrightarrow H(p,q)=-p^2+q^2}.
\end{equation}
In ~\cite{rosales-martin}, it was demonstrated that this function can be considered as the bounded Hamiltonian of a quantum system, thereupon obtaining self-consistency with number theory if, and only if, the cardinal of the factorization ensemble is equal to the dimension of the corresponding  Hilbert space. Here we also show a physical realization of the quantum device  supporting Eq. (\ref{invertedharmonic1}).

\bigskip
\color{prusia}{

\section{Formulation of the quantum conditions.}

} 
For the primes in $\mathbf{\mathfrak{F}}(j)$, one gets $q<q_m $ for some upper bound in the ensemble. Then, we are allowed to use quantum transformation theory  to build the finite and normalizable quantum amplitude of probability $\mathbf{\Psi}(q)$ for the stationary states described by the $q-$numbers and $\hat{p}\rightarrow i\partial_q$:
\begin{equation}\label{Schroedinger}
 \color{redmath}{\hat{H}(\hat{p},q)\mathbf{\Psi}(q)=E\mathbf{\Psi}(q)}.
\end{equation}
The constraints are $ \Psi(\sqrt{E})=0$, and  $\Psi(q_m)=0$.
The solutions are stationary waves,
\begin{equation}\label{solutions}
\color{redmath}{\mathbf{\Psi}(q)= q e^{-i\frac{q^2}{2}}\{\color{veneciano}{\mathit{F}}\color{oxford}{(\alpha,\tfrac{3}{2},iq^2)+d(E)\cdotp \color{air}{\mathit{U}}\color{oxford}{(\alpha,\tfrac{3}{2},iq^2)\}}}}.
\end{equation}
Here, $\color{veneciano}{\mathit{F}}\color{oxford}{(a,b,z)}$ and $\color{air}{\mathit{U}}\color{oxford}{(a,b,z)}$ denote the two linearly independent confluent hypergeometric functions, $\alpha=-\frac{iE}{4}+\frac{3}{4}$, and $d(E)$ is a suitable complex constant required to satisfy the constraints. The uniqueness of the solution of the Sturm-Liouville problem implies the quantization of $E$:
\begin{equation}\label{QuantumConditions}
\frac{\color{veneciano}{\mathit{F}}\color{oxford}{(\alpha, \tfrac{3}{2},iq_m^2)\color{air}{\mathit{U}}\color{oxford}{(\alpha,  \tfrac{3}{2},iE)}}}
 {\color{veneciano}{\mathit{F}}\color{oxford}{(\alpha, \tfrac{3}{2},iE)\color{air}{\mathit{U}}\color{oxford}{(\alpha,  \tfrac{3}{2},iq_m^2)}}}=1,
\end{equation}

If $E=o(1)$, i.e, for factors close to $\sqrt{N}$, one can solve Eq. (\ref{QuantumConditions}) as a series $E=1+\varepsilon(q_m)+o(\varepsilon^2)$.  After a straightforward,  albeit long, calculation we get (see Appendix)
\begin{equation}\label{epsilon1}
  \varepsilon(q_m)\approx\frac{1}{\log q_m}\{\tan\phi_0+\sin\phi_m\sec\phi_0\},
\end{equation}
where
\begin{equation}\label{phase}
\phi_m=q_m^2-\log q_m-\phi_0+\chi,
\end{equation}
$\phi_0\approx 1.11965$, is a universal constant and $\chi$  is an arbitrary phase that can be added to the solution since it is a periodic function of the coordinate upper bound $q_m$. Indeed,this is an equation that relates $E$ with the integers because by construction $q_m$ should be calculated for some prime in the ensemble. To see this, assume we can (classically) determine that $x$ is much larger than some known bound $B_{\mathcal{G}}$ (the index $\mathcal{G}$ comes from the word "gauge") and that, for  other numbers $N_k\in \mathbf{\mathfrak{F}}(j)$, the  bound $B_k$ is a  prime in the vicinity of $B_{\mathcal{G}}$. Then
\begin{equation}\label{ValuesinFj}
  \pi(B_k)=\pi(B_{\mathcal{G}})-k.
\end{equation}

We should consider  $q_{\mathcal{G}}\equiv q[\pi(B_{\mathcal{G}})]$ as the initialization of the algorithm that calculates the quantum spectrum and, therefore, we should prepare the state $\mathbf{\Psi}(q)$ with the boundary condition at $q_m(k)= q[\pi(B_k)]$, for $k$ integer.

In principle the election of the bound $\pi(B_{\mathcal{G}})$ is arbitrary but, as a matter of fact, since $\varepsilon\ll 1$, and $E=o(1)$, we intend to map values of $E$  that correspond only  to factors $o(\sqrt{N})$. Now, to initialize the quantum sieve,  first we should  declare algorithmically the search scenario, i.e., primes $B_{\mathcal{G}}\leq x\leq \sqrt{N}$. On the other hand, recall  that there are standard classical factorization algorithms, say, e.g., Pollard-$\rho$ ~\cite{Pomerance}, which are already quite efficient whenever the factor $x\lesssim \sqrt[4]{N}$. In $\mathbf{\mathfrak{F}}(j)$ we are not concerned on those cases so we could always consider bounds $B_{\mathcal{G}} \gg  \sqrt[4]{N}$.  These considerations taken into account, we will start the calculation of the spectrum of $\mathbf{\mathfrak{F}}(j)$ at
\begin{equation}\label{Ansatz}
B_{\mathcal{G}}=\wp\{ \nu \sqrt[3]{N}(\log \sqrt{N})^{\mathcal{G}}\}\ll \sqrt{N},
\end{equation}
for some power of the $\log\sqrt{N}$,  $\mathcal{G}=o(1)$. Here the arithmetic  function $\wp\{\cdotp\}$ calculates the closest prime to its argument and $\nu$ is some constant.  Computing the value of the variable $q_{\mathcal{G}}$ for this bound we get (see Appendix)
\begin{equation}\label{sizelength}
q_{\mathcal{G}} =3/8\nu^{-1} \frac{ \sqrt[6]{N}}{(\log\sqrt{N})^{\mathcal{G}}}.
\end{equation}
To simplify notation, we can take $\nu=3/8$. From this we are allowed to compute,  for $ \pi(B_k)$, the value of  $q_m(k)$. Using the prime number theorem we can develop it in terms of a series depending on the small parameter $\lambda= q_{\mathcal{G}}^2/\sqrt{N}$
\begin{equation}\label{series}
  q_m(k)\sim q_{\mathcal{G}} + 2/3\lambda k+o(\lambda^2k^2),
\end{equation}
meaning that $q_{\mathcal{G}}$ is essentially the size of the simulator.   Finally,  let us impose this boundary condition for $B_{\mathcal{G}}\in\mathbf{\mathfrak{F}}(j)$, upon selecting the arbitrary phase as
\begin{equation}\label{chi}
\chi(q_{\mathcal{G}})\equiv -q_{\mathcal{G}}^2+\log q_{\mathcal{G}}.
\end{equation}
Which, together with Eqs.~(\ref{series}), (\ref{epsilon1}) and (\ref{phase}) provides the quantization of the energies depending of two quantum numbers $\mathcal{G}$ and $k$. Discrete values of $\mathcal{G}$  are readily computed because in Eq. (\ref{sizelength}), $q_{\mathcal{G}}$ can be any of the discrete zeroes of the wave function $\mathbf{\Psi}(q)$, for $E=1$, according to our approximations.  The first quantum number is therefore related to the preparation of a state whose classical coordinate bound must be given by $q_{\mathcal{G}}$, while $k$ denotes every possible transition of the system to a different energy level from this initial state. This picture is entirely analogous to that of the Hydrogen atom,  where there are different series of spectroscopic transitions depending on the initial state (Rydberg formula). This explains the spectral behavior of the arithmetic function $E_k$ in Fig. \ref{fig:spectrum}.

Now, a  Taylor series of Eq. (\ref{epsilon1})  gives for the energies the more simplified form
\begin{equation}\label{solutionE}
E_k(\mathcal{G})\approx 1+\frac{k}{k_m}\frac{2\pi}{\log q_{\mathcal{G}}}  +o(k/k_m)^2,
\end{equation}
here, we denoted as $k_m=\frac{3}{2}\pi(\log\sqrt{N})^{3\mathcal{G}}$ the period of Eq. (\ref{epsilon1}) which corresponds to the number of stationary states of the simulator for the selected gauge $\mathcal{G}$. Remarkably, it only scales logarithmically with the number $N$, which is a similar result to Shor's for the quantum algorithm complexity of the integer factorization problem~\cite{shor}.

On the other hand, given the arbitrariness of the gauge, a kernel density estimation average can be determined. Then, the predicted spectrum from our quantum mechanical solution Eq. (\ref{solutionE}), that corresponds to the distribution of the primes  in $\mathbf{\mathfrak{F}} (j)$ can be compared versus de actual one. To this aim, in Fig. \ref{fig:j_var}  we have represented the density plots for the quantum mechanically predicted  pairs $(E_k,x_k)$ and  those obtained from counting the primes in $\mathbf{\mathfrak{F}}(j)$. The probability for a prime to be a factor is highest in the red area and lowest in the blue areas.  As expected from P\'{o}lya's conjecture, they have to be equal for $N \gg 1$. In our case, the calculation shows a remarkable agreement between both. Thus, quantum mechanics obtains in $\mathbf{\mathfrak{F}}(j)$ the apparently unpredictable jumps in the distribution of the primes, a feature that confirms again the relation between the physics of the quantum simulator and number theory. This result is additional to the prediction of the regular behavior of $\pi(x)$ in ~\cite{rosales-martin}.

To mimic the output of the simulator we have to make a Montecarlo calculation replacing, in Eq. (\ref{solutionE}), $\sqrt N$ by  $\sqrt{N^{'}}\in (\sqrt{N}-\log\sqrt{N},\sqrt{N}+\log\sqrt{N})$, i.e., other values in $\mathbf{\mathfrak{F}}(j)$ that contribute with slightly different  energy levels $E_k(\mathcal{G^{'}})$. We then calculate $x_k(E_k(\mathcal{G^{'}}))$ after the definition in Eq. (\ref{Energies}). Finally we plot, using the kernel density estimation average from these values, the  distribution of $o(\log\sqrt{N})^{3}$ random points with this metric.  Note that since $E_k=E(\pi(x_k),\pi(N_k/x_k))$, we need Riemann's Fourier expansion of $\pi(x)$ in terms of the $\zeta(s)$ function zeroes (see Appendix).
To elaborate Fig. \ref{fig:j_var} we used a truncated series of Riemann's zeroes to approximate $\pi(x)$.

Calculations,  for $N\approx 1.09693\times 10^{10}$,($j=10000$), $N\approx33.6412\times 10^{12}$, ($j=400000$) and  $N\approx 1.13673\times 10^{24}$, ($j=4\cdotp 10^{10}$)  are presented in Fig. \ref{fig:j_var}. The last number is the largest one that we can calculate using the prime tables available in {\cal Mathematica}$^\circledR$.  As expected, the number of $\zeta(s)$ zeroes required to calculate $x_k(E_k)$ is increasing. However, the increase is moderate, needing only 1000 zeroes for the largest factorization to correctly reproduce the features at this energy scale.  Note also that the precision of the result is better for very large $N$'s, a feature that supports our statistical approach for the quantum factorization problem.

Recall that, in order to compute the plots from quantum mechanics, only Eq. (\ref{solutionE}) and Riemann hypothesis have been used. Thus, these figures are descriptive examples of how quantum number theory provides the distribution of the primes below $\sqrt{N}$ with a precision only achievable if we knew the exact position and values of the pairs of primes in $\mathbf{\mathfrak{F}}(j)$.  According to Feynman's remark ~\cite{Feynman}, exact quantum probabilities can not be calculated with a classical computer. This corresponds to the fact that, even though $E$ can be interpreted as the arithmetic function in  Eq.~(\ref{Energies}), inverting $x(E)$ is not possible with complete accuracy given that it would require the inversion of an infinite series (with terms involving all the $\zeta(s)$ zeroes on the critical line). Notwithstanding with these considerations, we  were able obtain a good approximation to that probability with a truncated  Riemann's series.

It is important to realize that, in order to factorize the number $N$, one only has to reach values of $x$ using the probability distribution achieved quantum mechanically. Indeed, these graphs can be considered as the direct output of the energy measurements of the quantum simulator. A physical realization of such a simulator is the subject of the next section.

\begin{figure}[hbtp]
\centering
\includegraphics[width=0.471\textwidth]{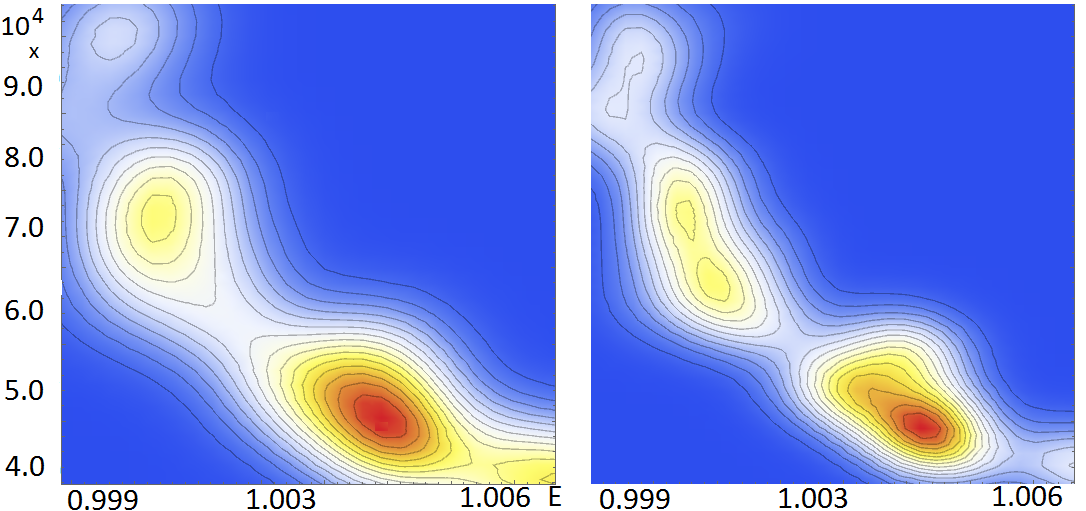}
\includegraphics[width=0.4863\textwidth]{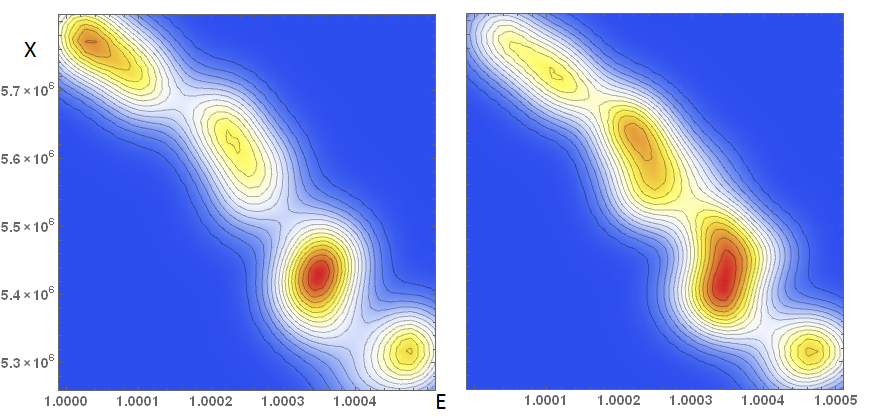}
\includegraphics[width=0.482\textwidth]{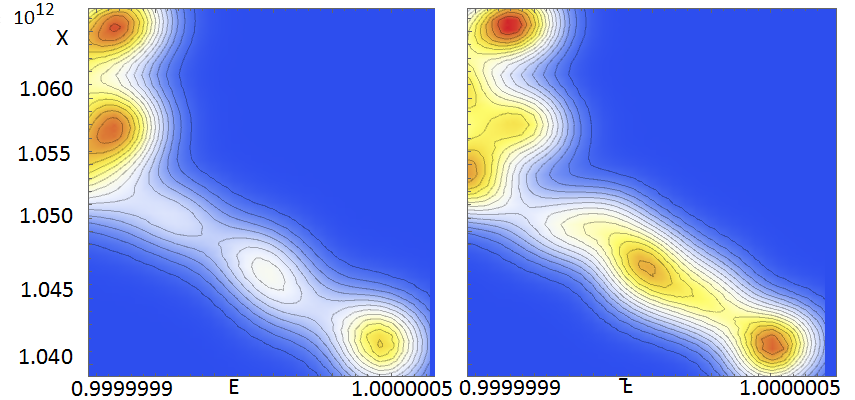}

\caption{\color{oxfordfootnote}{Comparative plot of the density distribution of values $(E_k,x_k)$  calculated quantum mechanically for the simulator (left column)  and counting exactly the primes (right column). For the quantum calculation, we have used a wide range of $j$ values :   $\mathbf{\mathfrak{F}}(10000)$,  $\mathbf{\mathfrak{F}}(400000)$ and an extreme value of $\mathbf{\mathfrak{F}}(4 \cdotp 10^{10})$(from top to bottom) versus the same figures computed counting the primes for the respective $j$. As expected, both distributions are similar.  It is to be noted that to achieve this precision, the number of $\zeta(s)$ function zeroes required for the truncated  series of $\pi(x)$ is reasonably small. The actual number of zeroes is different for each figure:  $100$, $300$ and  $1000$ (again from top to bottom). Due to the very high computational cost involved, the last classical calculation (bottom right) was not done using all primes.
}}
\label{fig:j_var}
\end{figure}

\bigskip
\color{prusia}{

\section {Physical realization of the quantum state: entangled particles in a Penning trap.}

}
The  simulator is a physical system. To see a possible experimental set up, first in Eq. (\ref{Schroedinger}) we make the substitutions
\begin{eqnarray}
\label{Transformation}
q\rightarrow\color{math}{ (\frac{M\omega_z}{2^{\frac{1}{2}}\hslash})^{\frac{1}{2}}} \color{oxford}{\varrho} , \; \nonumber \\
 E\rightarrow \color{math}{-\frac{2^{\frac{3}{2}}}{\hslash\omega_z}} \color{oxford}{ E^{'}}, \;\; \nonumber \\
\mathbf{\Psi}(q)\rightarrow \color{math}{\varrho^{\frac{1}{2}}}\color{oxford}{ \psi(\varrho).}\;\;
\end{eqnarray}
we get
\begin{eqnarray*}\label{Schroedinger_exact}
  \{\frac{\hslash^2}{2M}[-\frac{1}{\varrho}\partial_\varrho(\varrho \partial_\varrho)+
  \frac{(\frac{1}{2})^2}{\varrho^2}]-\frac{M\omega_z^2}{2}\frac{\varrho^{2}}{2}\}\psi=E^{'}\psi.
\end{eqnarray*}

Now, neglecting the term $\frac{\hslash^2}{2M}\frac{(\frac{1}{2})^2}{\varrho^2}\ll\frac{M\omega_z^2}{2}\frac{\varrho^{2}}{2}$, it provides, identifying $\varrho$ as a polar radial coordinate, very approximately for $\varrho\gg (\hslash/2^{\frac{1}{2}}M\omega_z)^{\frac{1}{2}}$
\begin{equation}\label{Schroedinger_approximate}
  -\frac{\hslash^2}{2M}\frac{1}{\varrho}\frac{\partial}{\partial\varrho}\varrho \frac{\partial\psi}{\partial\varrho}-\frac{M\omega_z^2}{4}\varrho^{2}\psi\approx E^{'}\psi.
\end{equation}
This is readily interpreted as the Schr\"{o}dinger equation satisfied by  a $l_z=0$ state with axial symmetry. The  classical limit must then correspond to a system which is confined, both radially and axially, in interaction with the potential energy $-\frac{M\omega_z^2}{4}\varrho^{2}$ which could be identified as an  electrostatic field. The form of this potential directly leads to identify Eq. (\ref{Schroedinger_approximate}) with the Hamiltonian  constraint corresponding to the physics of a Penning trap ~\cite{Penning}. In this system, the charged particles remain trapped radially by a magnetic field and axially by an electrostatic field which means that the particles must have also spin. Moreover, compatibility of the zero axial component of the angular momentum suggests that the system can be, for instance,  a $p$-wave  or a $s$-wave of two particles. These features might correspond to two different kind of Bell states made by two particles with mass $m_e=M/2$: \[|p_\pm\rangle= \frac{1}{\sqrt{2}}(|\color{spin2}{\uparrow}\color{spin1}{\uparrow}\color{oxford}{\rangle \pm }|\color{spin2}{\downarrow}\color{spin1}{\downarrow}\color{oxford}{\rangle)}\] or \[|s_\pm\rangle= \frac{1}{\sqrt{2}}(|\color{spin2}{\uparrow}\color{spin1}{\downarrow}\color{oxford}{\rangle \pm} |\color{spin2}{\downarrow}\color{spin1}{\uparrow}\color{oxford}{\rangle)}\] (recall that, if for  a $p$-wave,   $E^{'}\rightarrow E^{'}+g\hat{s}(e\hslash/m_e c) B$, g denotes the $g-$factor of the ion). Moreover, note that in Eq. (\ref{Schroedinger_approximate}) no electrostatic interaction appears and therefore the Coulomb energy must be added as a constant $e^2/x_0$, where $\pm x_0/2$ are the coordinates of the diametrally  opposed equilibrium positions for the two particles. The coordinate wave function is, thus, antisymmetric, which leads to  spin symmetric $p_+$ or $s_+$ possible configurations for half integer spin particles and to  spin antisymmetric $p_-$ or $s_-$ for integer spin ones. A way to experimentally build these kind of equilibrium Bell states was proposed in ~\cite{cirac} using, first, a rotational barrier to spatially separate the particles and then, in order to achieve entanglement, a resonant oscillating electric field is applied to drive the axial degree of freedom.  Once the Bell state is achieved, to get the interaction back to that in Eq. (\ref{Schroedinger_approximate}), the rotational barrier must be  adiabatically reduced to zero in time scale larger than the inverse of the Rabi frequency of the drive.  This procedure transfers a spatial antisymmetric configuration to the quantum state compatible with $l_z=0$.

The solution of the orbital part of Eq.~(\ref{Schroedinger_approximate})  (regardless the  asymmetric angular part that defines the line axis between the particles) is
\begin{eqnarray}\label{Schroedinger solution_approximate}
 \psi(\varrho)= \Re\{ e^{i\frac{\varrho^2}{2}}[\color{air}{\mathit{U}}\color{oxford}{(\beta,1,-i \varrho^2) +}  \\ \nonumber
 c_{E^{'}} \color{veneciano}{\mathcal{L}}\color{oxford}{(-\beta,-i \varrho^2)]\}}.
\end{eqnarray}
Again, $\color{air}{\mathit{U}}\color{oxford}{(a,b,c)}$ is the second confluent hypergeometric function, $\color{veneciano}{\mathcal{L}}\color{oxford}{(a,b)}$ is the generalized Laguerre function, $\beta=\frac{1}{2} - iE^{'}/(\sqrt{2}\hbar\omega_z)$ and  $c_{E^{'}}$ is calculated from the boundary condition $\psi(2/\omega_z\sqrt{-E^{'}/M})=0$.

We can check that  this wave function $\psi$  satisfies  the boundary conditions for the integrable probability density of the entangled state  (in the center of mass coordinate frame) at $\varrho\rightarrow 0$: \[\lim_{\varrho\rightarrow 0} \varrho \psi(\varrho)^2=0.\]   Moreover, in order to simulate the factorization conditions of Eq.~(\ref{Schroedinger}) we must additionally impose \[\lim_{E\rightarrow 1}\lim_{\varrho\rightarrow (E \hslash/(\sqrt{2}m_e\omega_z)^{\frac{1}{2}}} \varrho \psi(\varrho)^2=0.\]

Having these bounds, the rightness of the approximation can  be seen in Fig. \ref{fig:wavefunction} where we plotted the exact solution Eq.~(\ref{solutions}) and that from the Penning trap Hamiltonian with $l_z=0$. Therefore the solution provides correctly  all the zeroes of the exact wave function of the factoring simulator. In the figure we use $E\approx 1$.

\begin{figure}[hbtp]
\centering
\includegraphics[scale=0.55]{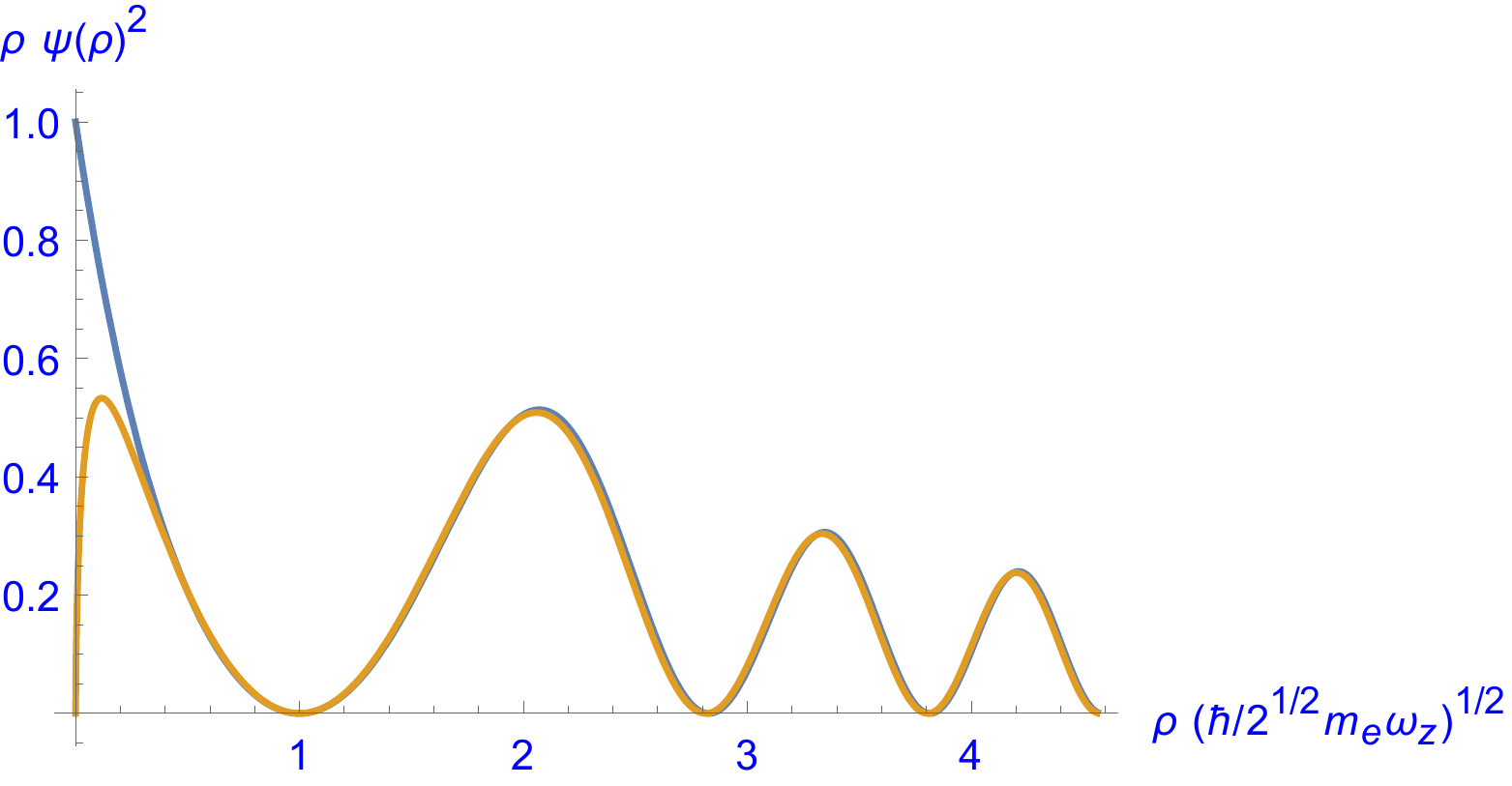}
\caption{ \color{oxfordfootnote}{Plot of the exact integrable probability density $d\mathcal{D}(\varrho,\psi)/ d\varrho =\varrho\psi(\varrho)^2$ for the factoring simulator Hamiltonian (in blue) versus the one calculated for the  Penning trap Hamiltonian (in orange) with the same boundary condition at $E=1$.  In order to become a physical realization of the factoring simulator both systems must have the same zeroes for $q\geq 1$ which is indeed the case.}}
\label{fig:wavefunction}
\end{figure}

Let us now describe in details how to program the Bell states in the Penning trap as the simulator that determines the spectrum of the factorization ensemble $\mathbf{\mathfrak{F}}(j)$. To this aim we should be able to determine the physical parameters of the trap for the numbers in this set, i.e., $\sqrt{N_k}\sim \sqrt{N}+\xi_k \log\sqrt{N}$ with $ \xi_k=o(1)$.

The constant cyclotron frequency is given by $\mathbf{\omega_c}= g\hat{s}  e/(m_e c) \mathbf{\mathcal{ B}}$. In a Penning trap, the classical motion is fully described by three frequencies namely, the cyclotron frequency $\omega_c$, the electrical oscillator $\omega_z$ and the magnetron frequency for the radially confined
motion $\omega_m$. Thus  Eq.~(\ref{Schroedinger_approximate}) represents the quantization of the magnetron motion.
Moreover, it holds that, for the trap to be operative, $
\omega_c\gg \omega_z\gg\omega_m$,  $\omega_{m}\approx \omega_z^2/2\omega^{'}_c$, with $\omega^{'}_c$ slightly smaller than the cyclotron frequency  and $\omega_c \geq \sqrt{2}\omega_z$.
Then, the $z-$motion decouples from the radial coordinate magnetron motion resulting in a simple harmonic oscillator of frequency $\omega_z$. This experimental set up confines the charges in a saddle point region $\varrho, z$, where $\varrho<\varrho_m\leq \varrho_0$, corresponding to the size $\varrho_0$ of the ring electrode. The quantized energy of the magnetron motion is exactly decoupled from the cyclotron motion  ~\cite{Penning}. It is given by,
\begin{eqnarray}\label{Energies_Penning_trap}
E^{'}=-\hslash  \omega_{m}(k+\frac{1}{2})
\Leftrightarrow E\approx \sqrt{2}\frac{\omega_z}{\omega^{'}_c}(k+\frac{1}{2}),
\end{eqnarray}
which corresponds to the linear dependence calculated for the isolated inverted harmonic oscillator, Eq.~(\ref{solutionE}), if and only if we impose
\begin{equation}\label{Penning Frequency_condition}
\frac{\omega_z}{\omega^{'}_c}=1/\log q_{\mathcal{G}}\cdotp \frac{\pi\sqrt{2}}{ k_m}.
\end{equation}

On the other hand, the size of the electrostatic  potential restricts the trajectories of the particles to the saddle point area of radius $\varrho_m$ say. Then $\psi(\varrho)$ will be non zero if $\varrho<\varrho_m$  and $\psi(\varrho)=0$ otherwise.  It reads
\begin{equation}\label{Penning_trap_Size}
        \varrho_m\approx (\frac{\hslash}{\sqrt{2}m_e\omega_z})^{\frac{1}{2}}q_{\mathcal{G}} \Leftrightarrow \omega_z=\frac{\hslash q_{\mathcal{G}}^2}{\sqrt{2} m_e\varrho_m^2}.
\end{equation}
Now from Eq.~(\ref{Penning Frequency_condition}), Eq.~(\ref{sizelength})  taken into account, one  finally gets, for the number that a technologically achievable trap is able to simulate,
\begin{eqnarray}\label{N}
\sqrt{N} \approx \frac{2^{\frac{3}{2}}}{3}\frac{ q_{\mathcal{G}}^3}{\log q_{\mathcal{G}}} \frac{\omega^{'}_c}{\omega_z}
\end{eqnarray}
which, since $\omega^{'}_c\gg\omega_z$, can  also be written as \[ \displaywidth=1.0\linewidth \sqrt{N} \sim \frac{q_{\mathcal{G}}}{\log q_{\mathcal{G}}}(\frac{4}{3} g\hat{s})[\frac{\pi\varrho_m^2 B}{h c/2e}].\]
Recall  that the term inside the square brackets is the quantum of the magnetic flux through the trap, $n$, implying that only discrete energy  levels are permitted (Landau levels). Therefore the number of operative qubits of the simulator is  exponentially large $2^n$. This physical set up is akin to the proposal in ~\cite{sierra} to experimentally determine Riemann's zeroes if  P\'{o}lya and Hilbert conjecture applies.

For electron traps  with $ \varrho_m\sim 3$ mm,  typical electrostatic fields yield to $q_{\mathcal{G}} \lesssim 10^2$ while $(\omega_c/\omega_z)\sim 10^3$, then  numbers up to $N\leq 10^{20}$ can be factorized with the quantum simulator.  Measuring the magnetron frequencies $ |E^{'}|/\hslash$,  gives $E= 2^{\frac{3}{2}}|(E^{'}-L g\hat{s} (e\hslash/m_e c)B)|/\hslash\omega_z$ ($L=0,1$ for the s-wave and p-wave, respectively). The only meaningful values being those with $E>1$.

The simulator operates as follows:  take, e.g., $q_{\mathcal{G}}=2.82765$, the first zero of the wave function  for $E=1$, then Eq.~(\ref{Penning_trap_Size}) fixes  $\omega_z$ and, in order to code the number we wanted to factorize in the simulator,  the magnetic field frequency is fine tuned with the help of Eq.~(\ref{N}). Then, upon selecting  new zeroes of the wave function $q_{\mathcal{G}}$, and re-scaling the field frequencies  of the trap accordingly, other  sets of energies of the simulator could be measured; this procedure, depending on the number of measurements, typically $o(\log\sqrt{N})^3$ (See Appendix), would yield to a detailed probabilistic density  pattern in a neighbourhood of each measured energy where the more probable values of $E_k$ will cumulate.  The result is a probabilistic quantum sieve for the more likely $x(E)$ factors in the ensemble. Given that linear jumps in $E$ ought to correspond to exponentially large ones in $x$, the simulator yields to an exponential speed up to find the factors, though recall that,  to finally exactly calculate  $x$ and $y$, a classical sieve would still be required provided the probability input for primes $o(\sqrt N)$.

\bigskip
\color{prusia}{

\section{Conclusion.}

}
In this paper we have shown how the quantization of the factoring hamiltonian Eq.~(\ref{invertedharmonic1}) leads  to a polynomial time algorithm that can be considered the analog equivalent of Shor's algorithm using a quantum, gate based, computer. This shows a  picture of the factorization problem as a quantum system whose energies provides insight on  the probability  of a given prime to be a  factor of the number $N$. This leads to an algorithm that bears no resemblance to any classical sieve.

For large $N$ the system can be solved to any precision, allowing to calculate the energy histogram of the system and compare it to the equivalent result in number theory. As such it can be seen as a test of the validity  of P\'{o}lya's conjecture related to Riemann's hypothesis.

We have calculated the statistical distribution of the energies as a function of one of the factors both, using the solution of the quantum simulator and resorting to classical number theory assuming P\'{o}lya's hypothesis, showing that they are indeed equivalent. These distributions cannot be explained from number theory alone, although they are readily understood when interpreted as the density of spectral lines of a quantum device as the one presented here.

Moreover, it is shown that the hamiltonian can be implemented as a physically accessible system consisting in preparing a Bell state for two particles in a Penning trap. Thus, in the end, the reason of the factoring exponential speedup of the simulator should be the coupling  of the entangled state with the quantized flux of the magnetic field.  Recall that,  apart from the $p_\pm$ and $s_\pm$ waves, other different symmetric spin states with more than two particles are compatible with the constraint $l_z=0$ in the Penning trap and that it  will likely  allow to  experimentally scaling the simulator even for much larger $N$'s.

\bigskip
\color{prusia}{

\section{Acknowledgements.}
}
We thank to Enrique Solano and Lucas Lamata for stimulating discussions and the suggestion that this proposal could be physically realizable and scalable with entangled ions in a Coulomb lattice configuration.

This work has been partially  supported by Comunidad Aut\'{o}noma de Madrid,  project Quantum Information Technologies Madrid, QUITEMAD+ S2013/ICE-2801.
and by project CVQuCo, Ministerio de Economía y
Competitividad, Spain, Project No. TEC2015-70406-R.
MINECO/FEDER UE.
}

\pagebreak
\appendixname

\bigskip
\color{oxfordapendix}{
{\bf Solution of the quantum conditions.}

\bigskip
We need to solve the equation
\begin{equation}\label{S}
  S(E,\rho) \equiv
\frac{\color{veneciano}{\mathit{F}}\color{oxford}{(\alpha, \tfrac{3}{2}, i\rho^2)\color{air}{\mathit{U}}\color{oxford}{(\alpha, \tfrac{3}{2},iE)}}}{ \color{veneciano}{\mathit{F}}\color{oxford}{(\alpha, \tfrac{3}{2},iE)\color{air}{\mathit{U}}\color{oxford}{(\alpha, \tfrac{3}{2},i\rho^2)}}}= 1,
\end{equation}
with $\rho = q_m^2\gg 1$ and $\alpha = \tfrac{3}{4} - i\tfrac{E}{4}$.

Let us work it out for $E = 1+\varepsilon(\rho)+o(\varepsilon(\rho)^2)$. Using Newton-Raphson at $E = 1$, denoting $S(\rho) = S(1,\rho)$, and $S^{'}(\rho) = \partial_E S(E,\rho)|_{E=1}$, we get up to first order
\begin{equation}\label{epsilon2}
\varepsilon(\rho) =
\frac{1 - S(\rho)}{ S^{'}(\rho)}.
\end{equation}
Now, in Eq.~(\ref{S}) take the Taylor expansions of $\color{air}{\mathit{U}}\color{oxford}{(\alpha,3/2,iE)}$ and  $\color{veneciano}{\mathit{F}}\color{oxford}{(\alpha,\tfrac{3}{2},iE)}$ near $E=1$;  also, for $\rho\gg 1$, we use \[\color{air}{\mathit{U}}\color{oxford}{(\alpha,\tfrac{3}{2},i\rho) \sim (i\rho)^{−\alpha},}\] and   \[\color{veneciano}{\mathit{F}}\color{oxford}{(\alpha,\tfrac{3}{2},i\rho)\sim  \tfrac{\sqrt{\pi}}{ |\Gamma(\alpha)|\rho^{3/4}} \cos(\arg \Gamma(\alpha) + \tfrac{3\pi}{ 8} + \tfrac{E}{4} \log \rho - \tfrac{\rho}{2})},\] where we further approximate $\arg \Gamma(\alpha)$ and  $ |\Gamma(\alpha)|$ by the Taylor series near $E=1$.  Provided with these formulas we get
\begin{equation}\label{S0}
 S(\rho) = e^{i\vartheta(\rho)} \cos \phi_0 \sec\varphi(\rho),
\end{equation}
with $\phi_0 \approx 1.11965$, $\vartheta(\rho) = \varphi(\rho) - \phi_0$, and $\varphi(\rho) = 1/4\log\rho - \rho/2 + C$. Here $C$ is an arbitrary constant that can be added since the function $S(E,\rho)$ is periodic. Also we obtain for $S^{'}(\rho)$
\begin{equation}\label{S1}
  S^{'}(\rho) = \frac{S(\rho)}{4} A(\rho),
\end{equation}
with $A(\rho) \sim (\tan\phi(\rho) + i)\log\rho$, asymptotically for $\rho \gg 1$. Lastly, feeding Eqs.~(\ref{S0}) and (\ref{S1}) in Eq.~(\ref{epsilon2}), we get after some additional algebra and re-grouping all the phase constants in $\chi$
\begin{equation}\label{solution}
\varepsilon(q_m) \sim
\frac{1}{ \log q_m}\{\tan\phi_0 + \sin(q_m^2 -\log q_m -  \phi_0+\chi) \sec\phi_0\},
\end{equation}
which coincides with Eq.~(\ref{epsilon1}) in the main text.

\bigskip
{\bf Calculation of the logarithmic number of states of the simulator.}

\bigskip
We are assuming $E=o(1)$, then $x=o(\sqrt{N})$; since $N=xy$, there exists a bound $B\ll x$. We select the "Ansatz" \[B_{\mathcal{G}}=\wp\{3/8\sqrt[3]{N}(\log \sqrt{N})^{\mathcal{G}}\},\]  for some constant $\mathcal{G}=o(1)$ (we denote as $\wp\{t\}$ the function that gets the closest prime of its argument). Now, there exists $N(B_{\mathcal{G}})$ in the ensemble, so that we calculate the co-prime \[A_{\mathcal{G}}=\frac{N(B_{\mathcal{G}})}{B_{\mathcal{G}}}= \wp\{8/3 N^{\tfrac{2}{3}}(\log\sqrt{N})^{-\mathcal{G}}\}.\] Thus, the prime number theorem yields to
\begin{align*}
 \pi(B_{\mathcal{G}})\sim \tfrac{9}{16} \sqrt[3]{N}(\log\sqrt{N})^{\mathcal{G}-1} ,\\
 \pi(A_{\mathcal{G}})\sim 2 N^{\tfrac{2}{3}}(\log\sqrt{N})^{-\mathcal{G}-1}.
\end{align*}
With those bounds, the calculation should only be valid for energies of the simulator such that $E\lesssim E(\mathcal{G})=\pi(B_{\mathcal{G}})\pi(A_{\mathcal{G}})/j^2 \sim  9/8$.

We need to compute other bounds $B_k$ for  $N_k=x_k y _k \in \mathbf{\mathfrak{F}}(j)$; if $N_k=o(N)$ let us write a proximity relationship for the corresponding $B_k$: \[\pi(B_k)=\pi(B_{\mathcal{G}})-k.\] Analogously there are others $N(B_k)$ in the ensemble and thus we calculate $A_k=\frac{N(B_k)}{B_k}$ that obtains
\[
\pi(A_k)\sim \pi(A_{\mathcal{G}})+\frac{4}{3} k \sqrt[3]{N}(\log\sqrt{N})^{-2\mathcal{G}-1}.
\]
Now, since $\pi(A_k)\gg \pi(B_k)$, we can make the approximation $q_m(k)\sim \frac{1}{2j}\pi(A_k)$, therefore,
\begin{equation}\label{qk}
  q_m(k)\sim \frac{\sqrt[6]{N}}{(\log\sqrt{N})^{\mathcal{G}}}+(2/3k) /\sqrt[6]{N}(\log\sqrt{N})^{-2\mathcal{G}}.
\end{equation}
Then, writing \[q_{\mathcal{G}}= \frac{\sqrt[6]{N}}{(\log\sqrt{N})^{\mathcal{G}}},\]
we  get
\begin{equation}\label{qk1}
q_m(k)\sim q_{\mathcal{G}}+(2/3k) q_{\mathcal{G}}^2/\sqrt{N}.
\end{equation}
Finally, for the phase in Eq.~(\ref{phase}) of the main text, if we redefine the arbitrary constant as
\begin{align*}
 \chi(q_{\mathcal{G}}) = - q_{\mathcal{G}}^2 +\log q_{\mathcal{G}},
\end{align*}
we get, using Eq.~(\ref{qk1}),
\begin{equation}\label{phim}
q_m(k)^2-q_{\mathcal{G}}^2\sim \frac{2\pi k}{k_m},
\end{equation}
where the period of $\varepsilon(q_m)$, i.e.,  the number of possible energy states of the simulator, only scales polynomially with the number of digits of  $N$
\begin{equation}\label{km}
  k_m=\frac{3\pi}{2}(\log \sqrt{N})^{3\mathcal{G}}.
\end{equation}

\bigskip
{\bf Kernel density estimation average.}

\bigskip
From the histogram of the computed values $E_k(\mathcal{G^{'}})$, the mixture kernel density function consists of a selected number of kernel densities with equally weighted coeﬃcients, $\upsilon_{k,\mathcal{G^{'}}}$ ~\cite{KDE}. It allows to compute each of the energy quantum levels simply as
\begin{eqnarray*}\label{statisticalaverageofE}
  <E_k>=\sum_{\mathcal{G^{'}}}\upsilon_{k,\mathcal{G^{'}}}E_{k}(\mathcal{G^{'}}).
\end{eqnarray*}
Moreover we can estimate the inferred width for each of the energy levels
\begin{eqnarray*}\label{widthofE}
  \sigma_k^2(E)=\sum_{\mathcal{G}}\upsilon_{k,\mathcal{G}}E_{k}(\mathcal{G})^2-<E_k>^2.
\end{eqnarray*}

\bigskip
{\bf Numerical solutions of $x(E)$.}

\bigskip
They could be obtained upon inverting  the  approximate Fourier expansion of $\pi(x)$, expressed as the truncated finite series depending on the Riemann zeta function  zeroes $\rho_k=\frac{1}{2}+iT_k$, given that $  \pi(x)\approx R(x)- \sum_k^T R(x^{\rho_k}) $, where $R(x)$ is the Riemann prime counting function, denoting $\eta_T(x)=-\sum_k^T R(x^{\rho_k})/R(x)$,
a solution of the equation
 \begin{align*}
E\sim \frac{R(x)R(\frac{N}{x})}{j^2}(1+\eta_T(x))(1+\eta_T(\frac{N}{x})),
 \end{align*}
 must be obtained. Depending on the cutoff $T$, this procedure will  give, in principle, $x(E)$ with any desired exactitude.
}

\end{document}